\newcommand\pa{\partial}

\newcommand{\beque}{\begin{equation*}}
\newcommand{\eeq}{\end{equation}}
\newcommand{\beq}{\begin{equation}}
\newcommand{\eeque}{\end{equation*}}
\newcommand{\beqnl}{\begin{eqnarray}}
\newcommand{\eeqna}{\end{eqnarray*}}
\newcommand{\beqna}{\begin{eqnarray*}}
\newcommand{\eeqnl}{\end{eqnarray}}

\documentclass[aps,prd,superscriptaddress]{revtex4}
\usepackage{amsfonts}
\usepackage{amsthm}
\usepackage{amsbsy}
\usepackage{amssymb}
\usepackage{amsmath}
\usepackage{graphicx}

\begin{document}

\date{\today}

\title{Tunneling approach and thermality in dispersive models of
 analogue gravity}

%\author{F.~Belgiorno$^1$, S.L.~Cacciatori$^{2,3}$, F.~Dalla~Piazza$^4$}
%
%\address{$^1$ Dipartimento di Matematica, Politecnico di Milano, Piazza Leonardo 32, IT-20133
%Milano, Italy\\
%$^2$ Department of Science and High Technology, Universit\`a dell'Insubria, Via Valleggio 11, IT-22100 Como, Italy\\
%$^3$ INFN sezione di Milano, via Celoria 16, IT-20133 Milano, Italy\\
%$^4$ Universit\`a ``La Sapienza'', Dipartimento di Matematica, Piazzale A. Moro 2, I-00185, Roma,   Italy}

\author{F.~Belgiorno}
\affiliation{Dipartimento di Matematica, Politecnico di Milano, Piazza Leonardo 32, IT-20133 Milano, Italy and INdAM-GNFM}
\email{francesco.belgiorno@polimi.it}
\author{S.L.~Cacciatori} 
\affiliation{Dipartimento di Scienza e Alta Tecnologia, Universit\`a dell'Insubria, Via Valleggio 11, IT-22100 Como, Italy and INFN sezione di Milano, via Celoria 16, IT-20133 Milano, Italy}
\email{sergio.cacciatori@uninsubria.it}
\author{F.~Dalla~Piazza}
\affiliation{Universit\`a ``La Sapienza'', Dipartimento di Matematica, Piazzale A. Moro 2, I-00185, Roma,   Italy}
\email{dallapiazza@mat.uniroma1.it, f.dallapiazza@gmail.com}

\begin{abstract}

We set up a tunneling approach to the analogue Hawking effect 
in the case of models of analogue gravity which are 
affected by dispersive effects. An effective Schroedinger-like equation for the basic 
scattering phenomenon $IN\to P+N^\ast$, where $IN$ is the incident mode, $P$ is the 
positive norm reflected mode, and $N^\ast$ is the negative norm one, signalling 
particle creation, is derived, aimed to an approximate description of the phenomenon. 
%Group velocity 
Horizons and barrier penetration play manifestly a key-role 
in giving rise to pair-creation. The non-dispersive limit is also correctly recovered. 
Drawbacks of the model are also pointed out and a possible solution ad hoc is suggested.

\end{abstract}

%\pacs{04.70.Dy,04.70.-s,03.65.Pm}
%\pacs{04.62.+v, 04.70.Dy}
\maketitle

\section{Introduction}
\label{intro}

Since the original work by Hawking, 
the fundamental mechanism leading to Hawking radiation has been indicated as quantum mechanical 
tunneling of a particle through the black hole horizon \cite{hawrad}. 
In particular, in the geometrical optics approximation, which is justified 
by the fact that near the horizon particle energies are severely boosted, because of an 
hard blueshift, a classically forbidden tunneling of a particle from inside the horizon 
to outside can occur, according to still different mechanisms. 
Together with other approaches, so-called tunneling methods appeared since the initial period 
in black hole evaporation calculations, and we limit 
ourselves to quote only some seminal papers and a fine review 
\cite{damour-ruffini,padmanabhan,visser-essentials,parikh-prl,vanzo}.  It is also worth mentioning 
that in the Parikh-Wilczek approach to the Hawking effect 
tunneling through the horizon of a particle can happen 
because of a quite unexpected mechanism, where the tunneling particle sets up the barrier by 
energy conservation, as nicely described by Parikh \cite{parikh-secret}. In this method, 
as well as in the Hamilton-Jacobi one, coordinates well-defined 
on the horizon are to be preferable because they are not involved in subtle problems of analytical 
continuation (see e.g. \cite{vanzo}). The spectrum results to be non-strictly 
thermal because of the backreaction which takes place explicitly in the method, and which is 
fundamental for its feasibility \cite{parikh-prl,parikh-secret}.\\

As well-known, in the dispersive case geometrical concepts 
like metric and so on are nearly lost (it is true that one could appeal to concept 
like `rainbow metrics' and so on, but apparently there is no gain in pursuing this route in 
analogue gravity). Still, geometrical optics, and in particular geometrical optics tools which 
occur in the aforementioned methods, can be fruitfully adopted also in this analogue gravity 
framework. We shall discuss this topic with a particular reference to the optical black hole case. 
Since the first explorations in the field \cite{unruh-dispersion,corley-j}, thermality appeared to be 
preserved in analogous black holes, despite the presence of dispersion. See also \cite{barcelo-rev}. Moreover, 
group velocity horizons play the role 
of black hole horizons in the dispersive case. We start by analyzing the most natural concept for 
replacing a geometrical (kinematical) horizon, i.e. the would-be event horizon of the non-dispersive 
analogue case: the group velocity horizon. A group horizon is defined as the geometrical locus where 
the group velocity of the particle falls to zero in the comoving frame, or to the locus where the group 
velocity reaches the limit velocity of the model 
(light velocity in the optical case or sound velocity in fluid-like models) 
in the lab frame. There is a double question which is to be posed: 
is the group horizon responsible of the 
particle creation tout-court, or it is more simply responsible of making the phenomenon of pair-creation 
much more efficient? Moreover, in the latter hypothesis, which is the locus deputate to be responsible 
of particle creation? These questions are basic questions, but we must also underline a substantial 
fact: in the process of rebounce of the wave, which has been identified as a unavoidable ingredient in the 
pair-creation process in analogous models, there occurs a somehow non negligible region where the 
wave is enormously distorted. If the process is diabatic, particle creation occurs, and wave distortion is 
even bigger. The point is that, in the aforementioned region, reasonable doubts about the feasibility 
of concepts like group horizons maybe could be questioned, and in any case it would be hard to make a 
sharp distinction in the dynamics of the wave between group horizon position and phase horizon and also 
non-dispersive horizon. We try in the following to delve into the above questions, in the attempt to gain answers 
in an analytical way. In this sense, 
it is to be noted that, only a few analytical calculations appeared 
\cite{corley,himemoto,saida,Schutzhold-Unruh,unruh-s,coutant-analytic,Leonhardt-Robertson,Coutant-prd,Coutant-und,Un-schu,petev-prl,Coutant-thick}, 
but maybe no analytical definitive calculation, showing in a clear way and in a model independent way the key-role 
of group velocity horizons in the analogue Hawking effect in general dispersive media, has been provided.  In the 
case of fluid-like models, we remark that papers \cite{coutant-analytic,Coutant-thick} are very general in their 
discussion. Still, calculations are considerably (but maybe unavoidably) involved. See also \cite{belgiorno-hawking}, 
where nonperturbative analytical calculations are performed for the dielectric black hole case in the framework of the Hopfield model.\\

We propose an interesting model as an approximation and as a possible step to fill this gap. 
In particular, we consider a phenomenological model for optical dispersion 
and show that Hawking-like effect arises as tunneling by antiparticle states, 
as in the gravitational case \cite{damour-ruffini},  provided that 
a group horizon exists, through which a particle with energy $\omega$ can travel. 
Still, in a naive approach, 
the numerical coefficient is not the right one, in the sense that, even in the limit of negligible 
dispersion, it is seemingly not recovered the correct temperature, just for a numerical coefficient.\\
The latter problem could be considered to arise because of the nature of our approximation, 
and so it could appear that it could be considered as compatible, and even tolerable. Nevertheless,  
we think that the limit of negligible dispersion should be correctly reproduced, and then we 
investigate if a different expansion point, different form the group horizon, could give us a 
better answer. Quite surprisingly, provided that a suitable regularization is introduced, 
the right answer with the right coefficient is obtained at the horizon one finds in the 
non-dispersive situation, which is called geometrical horizon henceforth.  
This would be compatible with the idea that a weak dispersion 
effect is not and cannot be responsible of a drastic change in the nature of the process, 
and that the model we propose could be an interesting bridge between the non dispersive world 
and the dispersive one (where a weak dispersion is taken into account). Indeed, in the limit 
of negligible dispersion, the usual tunneling picture result of non-dispersive and astrophysical 
black holes is recovered. The method itself, in this sense, could belong to the long list of methods 
allowing to recover Hawking radiation in the non-dispersive case. 
As a matter of facts,  the limit as dispersive 
effects vanish (as $B\to 0^+$ in the optical model in the Cauchy approximation, and as $\Lambda\to \infty$ 
in the Coutant-Parentani-Finazzi model) is a singular limit involving both vanishing quantities and 
diverging ones, and has to be handled with care. 
Furthermore, in the same limit, the group horizon is shifted to coincide 
with the geometrical horizon of the non-dispersive model, so it needs itself a 
regularization procedure which sensibly modifies the way the nondispersive limit is approached. 
We find again the correct result.\\ 
There is still a problem. Even if the non-dispersive limit is correct, as dispersive effects arise and the 
expansion is made around the group horizon for nonzero $B$ (or finite $\Lambda$), a different 
temperature is found: a different coefficient appears, which corresponds to an enhancement of the 
temperature for a factor $3/2$. So, if we want to identify the group horizon as the locus where 
pairs are created, we find a discontinuous behaviour.\\ 
An alternative locus could be the geometrical horizon, again for $B>0$ ($\Lambda<\infty$), 
which could be considered as a better locus, to some extent: 
indeed, it would ensure that there is a continuous behaviour of the temperature in the non-dispersive 
limit, and, moreover, there would be a better coherence of the whole picture of analogue 
gravity. Last, but not least, there would be a better agreement both with experimental data and with 
numerical simulations (none of which seems to reveal a different temperature with respect to the 
one calculated in the non-dispersive case; in particular, the factor $3/2$ does not appear to 
be viable). But there is a discontinuous behavior which has to be taken into account, and 
which is even worse than in the case of the group horizon, as for non-zero dispersive effects 
there is no thermal behaviour at the geometrical horizon.\\
Even if the dichotomy between the behaviour of the relevant tunneling amplitude at finite dispersive effects 
and in the limit as dispersion vanishes 
can be considered as a serious drawback of our model, we point out that its 
simplicity and its being very near to the correct answers in the dispersive case, and its providing 
a further path, quite unusual, to find out the well-known non-dispersive ones (which are reachable 
by means of several methods), make it an interesting tool for discussing the dispersive case and 
to find a common method for discussing both cases. Its limits are also evident: the discontinuity 
in the temperature one finds if one focuses on the group horizon or on the geometrical horizon, 
is the major one. It would be possible to reconcile all the above discussion 
only if it were true that a divergence would occur at the group horizon also when dispersion is present. 
But it is not what emerges from the present analysis.\\
A further possible solution of the aforementioned problems, which allows to find both the correct limit 
as dispersive effects vanish and continuity in the behavior of the temperature in the same limit, 
without invoking a special prescription for the limit itself, is provided. Under hypotheses which 
are seemingly interesting, particle production could be associated with a further point, which we call 
`inner horizon', falling beyond the geometrical horizon and converging to it in the non-dispersive limit. 
This proposal has admittedly the drawback to be constructed ad hoc, 
in order to find out thermality up to corrections which vanish in the limit of negligible dispersion.\\
Our paper plan is the following: in sec. \ref{group-hor} 
we first discuss the problem of group velocity horizon, following and 
also modifying techniques adopted in fluid mechanics, and involving a suitable expansion of the 
dispersion relation around the group horizon, to be discussed in subsec. \ref{expansion-gh}. 
Then, in sec. \ref{expansion-geom} we consider a more radical modification of 
such an expansion, by drifting our attention into the non-dispersive horizon point. A link between the 
two expansions in the limit as dispersive effects vanish is also provided. In sec. \ref{disc-temp}, 
the discontinuous behavior 
of the temperature expression is pointed out, with a possible solution constructed ad hoc 
which is the argument of sec. \ref{inn-hor}. 
A final discussion appears in sec. \ref{concl}. For the sake of completeness, in appendix 
\ref{gh-gr} we show how concepts of geometrical optics can be applied also to the 
general relativistic case. In appendix \ref{pw-geomopt}, we apply a straightforward variant of the 
Parikh-Wilczek method to the case of analogue black holes, which displays thermality 
in the non-dispersive case.

\section{Group Horizon and Group Horizon Expansion} 
\label{group-hor}
 
Let $G$ stay for the dispersion relation in a 2D case, 
in a static situation where no explicit dependence on time $t$ occurs: 
this happens e.g. in the pulse reference frame in dispersive media, and one has
\beq
G(\omega,k,x)=0.
\label{disp-rel}
\eeq
$\omega$ is constant in the given framework. By solving e.g. for $k$, one obtains a codimension 1 
submanifold. 
The group velocity is given by 
\beq
v_g=\frac{dx}{dt}%=\frac{d\omega}{dk}
=\frac{\partial_k G}{\partial_\omega G}
\label{group-v}
\eeq
and a group horizon occurs for 
\beq
\partial_k G=0.
\label{group-h}
\eeq
Then, putting the above equation in a system with the dispersion relation, one obtains a codimension 2 
submanifold (up to exceptional configurations, which are not considered herein). In particular, we 
get solutions $x_{GH}$ which depend parametrically on $\omega$.

\subsection{Eikonal equation and group horizon}

In optical systems, 
we can generate a group horizon by means of a traveling perturbation of refractive index 
induced in a nonlinear dielectric medium by the Kerr effect \cite{leonhardt,rubino-njp,belgiorno-prl,
belgiorno-prd,petev-prl,finazzi-pra,finazzi-prl}. In this case, it is suitable 
(but not strictly necessary) to adopt the comoving frame of reference of the dielectric 
perturbation (which is assumed to be moving, within a good approximation, with constant 
velocity). In this framework, a seed pulse meets a group horizon when it falls to zero 
velocity, and this happens when it is traveling through the pulse signal. Needless to say, dispersion 
makes not so automatic to get a group horizon in the above sense, and suitable 
conditions have to be implemented. What follows is meant to complement the perturbative analysis 
which was carried out in \cite{petev-prl}.\\ 
For simplicity we take into account the case of the Cauchy approximation, which 
holds for frequencies much lower than the resonance one in the case of a single-resonance model. 
In particular, it holds $n(\omega_{lab})=n_0+B \omega_{lab}^2$, where $n_0$ does not depend on the 
lab frequency $\omega_{lab}$.  
We limit ourselves to considering only the branch which is involved with group horizons. 
Then, in presence of the Kerr effect, we have in the comoving frame \cite{belgiorno-prd}
\beq
G=0 \Longleftrightarrow (\omega + v k) (n(x)+B \gamma^2 (\omega+ v k)^2) - c k -\frac{v}{c} \omega=0.
\eeq
It is also useful to rewrite it as follows:
\beq
B \gamma^2 (\omega+ v k)^3 -\left( \frac{c}{v}-n(x) \right)(\omega+ v k)+\frac{c}{v} \frac{\omega}{\gamma^2}=0.
\label{di-cau}
\eeq
We take into account that $\omega$ is a variable separation constant in the comoving frame, so we can 
assume to solve the dispersion relation in $k$, obtaining $k_b = g_b (\omega,x)$, where $b$ labels the 
different branches. The cubic equation can be solved by means of Cardano's formulas. We are not interested 
in the explicit expression, which would be involved, but we are interested in the expression for the 
group horizon (if any), which is obtained by solving the system
\beqnl
G&=&0,\\
\pa_k G&=&0.
\eeqnl
As to the latter equation, we obtain
\beq
\pa_k G=0 \Longleftrightarrow 3 B \gamma^2 v (\omega + v k)^2-v \left( \frac{c}{v}-n(x) \right)=0,
\label{gh-cau-eq}
\eeq
which can be solved explicitly:
\beq
(\omega + v k) = \pm \left( \frac{\frac{c}{v}-n(x)}{3 B \gamma^2}\right)^{1/2}.
\eeq
By substitution of the positive root in $G=0$, as we mean to get the group horizon for positive norm 
waves (see the following section), we obtain an equation for $n(x)$ which 
allows us to find out explicitly the group horizon:
\beq
\frac{c}{v}-n(x) = 3 B \gamma^2 \left( \frac{1}{2 B \gamma^4} \frac{c}{v} \right)^{2/3} \omega^{2/3}=: \zeta_B \omega^{2/3},
\label{gh-cau-explicit}
\eeq
where $\zeta_B \propto B^{1/3}$. We also find 
\beq
(\omega + v k) |_0=\left(\frac{1}{2 B \gamma^4} \frac{c}{v}\right)^{1/3} \omega^{1/3},
\eeq
where all quantities are considered at the group horizon.\\
It can be noted that, as $B\to 0^+$, the group horizon converges to the 
geometrical horizon of the non-dispersive model: 
\beq
\lim_{B\to 0^+} x_{GH}(\omega)= x_{geom},
\label{x-ghgeom}
\eeq
where $x_{geom}$ is such that $n(x_{geom})=\frac{c}{v}$ \cite{belgiorno-prd}.\\ 
For example, in the case of a Gaussian pulse with $n(x)=n_0+\eta \exp \left(-\frac{x^2}{\sigma^2}\right)$, 
one finds
\beq
x_{GH}(\omega)=\pm \sqrt{2} \sigma \left[ -\log \left( \frac{1}{\eta} \left(\frac{c}{v}-n_0 -\zeta_B \omega^{2/3} \right)\right)
\right]. 
\label{x-gh}
\eeq
The expression (\ref{x-gh}) shows explicitly that the group horizon (if allowed) depends on $\omega$, and that for low 
values of $\omega$ it is almost indistiguishable from the horizon one finds in the non-dispersive model. So, at least 
in that region of frequencies, the link with the non-dispersive model is quite strong. To be more precise, 
the geometrical horizon of the non-dispersive case is replaced, in presence of dispersion, by a 1-parameter 
family of group horizons, with parameter $\omega$. If, as in the Cauchy approximation, we have $\omega 
\in (0,\omega_{max}]$, then this family runs from $x_{gh} (\omega_{max})<x_{geom}$ up to $x_{geom}$: 
$x_{gh}\in [x_{gh} (\omega_{max}),x_{geom})$. A different situation occurs when there exists a 
$\omega_{min}>0$, which is such that the geometrical horizon is never approached 
(cf. \cite{finazzi-pra,finazzi-prl}). 
One could even explore 
what happens for $\omega$=0. It is easy to show that (\ref{di-cau}) and (\ref{gh-cau-eq}) require both 
$k=0$ and $n=\frac{c}{v}$, the latter condition being the horizon condition of the nondispersive model. 
So we have undulation together with the nondispersive horizon condition. Still, there is a problem there, because of 
the vanishing of $\pa_k^2 G$ at the turning point, which would require the analysis of higher order contributions.\\
We could as well consider the case 
of a incompressible fluid, where the phenomenon can be classically described as blocking of 
waves \cite{weinfurtner-prl,rousseaux-book,rousseaux-njp}. 
Analogously, in BEC \cite{carusotto-bec,macher-bec} and in other materials, where horizons of this kind can be as well 
generated. See also the following section.

\subsection{Approximation near the turning point TP: the quantum case and the Hawking radiation}
\label{expansion-gh}

Let us consider the tunneling of a particle from inside a group velocity horizon. We start by recalling that  
the presence of a turning point, to be identified with a group horizon, is known to be a problem for 
the eikonal approximation. As a matter of facts, 
we are considering wave-like phenomena in which waves are strongly distorted at the rebounce on the dielectric perturbation 
(comoving frame). This implies that any concept like group velocity and ideas like the precise individuation of points is to be handled with 
care, having in mind the implicit limits in their use. Anyway, there is a possibility to provide an useful modification for the 
eikonal equation in the comoving frame, such that near the turning point a wave equation still holds true, allowing a better 
match between eikonal solutions and the presence of turning points, in the same spirit one finds in WKB approximation the 
Airy equation near the turning points of the potential in the nonrelativistic Schroedinger equation. We follow \cite{Peregrine-Smith}. 
See also the discussion in \cite{chaline}. We stress that the original studies are concerned only with a 
classical level analysis of wave propagation, and the equation derived is aimed only to provide  the amplitude of the wave equation 
\cite{Peregrine-Smith}. Furthermore, we do not match solutions with asymptotic ones, but simply we calculate 
the transmission coefficient in the aforementioned approximation. In other terms, we consider the pair creation phenomenon locally, 
in agreement with tunneling approach for black holes.\\
A similar expansion at the turning point, in the case of the 
elettromagnetic field in plasmas physics has been proposed e.g. in \cite{fuchs-ko-bers,fuchs-bers-harten}. 
Herein, our aim is to consider the same approximation for the dispersion relation at the 
turning point, with a qualifying difference: when we substitute for operators, we keep trace 
of the mixed term $\pa_x \pa_k G$ and we impose to get an equation involving an hermitian wave operator. 
Moreover, we mean to adopt the variable separation ansatz for the wave equation $\Psi (x,t)= 
e^{-i \omega t} \Phi (x)$, which is justified because of the static nature of the comoving frame 
(no explicit time-dependence of the refractive index).\\
The wave equation to be considered is 
\beq
G(\omega,k,x)\Psi (x,t)=0,
\eeq
where $\omega=-i\partial_t$ and $k=-i\partial_x$ are considered as operators in space-time variables. 
In the original approach, $\Psi (x,t)= b e^{i\eta}$, where $b$ is a slowly varying amplitude and 
$\eta$ is a rapidly varying phase \cite{Peregrine-Smith}. \\ 
We consider $\omega$ as a variable separation constant, due to the fact that $\omega$ is conserved 
in our static framework; terms up to the second order near the turning point $x=x_0$ such that $(\pa_k G)|_0 =0$ are: 
\beqnl
G&\sim& G(\omega,k_0,x_0)+(\pa_k G)|_0 (k-k_0)+ (\pa_x G)|_0 (x-x_0)\cr 
&+&
\frac{1}{2} (\pa^2_k G)|_0 (k-k_0)^2 +(\pa_x \pa_k G)|_0 (x-x_0)(k-k_0)+
\frac{1}{2} (\pa^2_x G)|_0 (x-x_0)^2.
\label{exp-G}
\eeqnl
In this expansion, we replace $k-k_0 \mapsto -i \pa_x$ and, moreover, in view of our 
aim of obtaining a Schroedinger-like quantum mechanical equation, we symmetrize the 
term proportional to $(\pa_x \pa_k G)|_0$ (Weyl symmetrization rule): we get
\beq
\left( -\frac{1}{2} (\pa^2_k G)|_0 \pa_x^2 -i (\pa_x \pa_k G)|_0 \frac{1}{2} ((x-x_0) (-i\pa_x)+(-i\pa_x) (x-x_0)) +
(\pa_x G)|_0 (x-x_0)+\frac{1}{2} (\pa^2_x G)|_0 (x-x_0)^2\right) \Phi(x)=0.
\eeq
Then we obtain an equation of the form 
\beq
\left(\pa_x^2 +2 i b_0 (x-x_0) \pa_x +(i b_0-2 c_0 (x-x_0)-d_0 (x-x_0)^2) \right) \Phi (x)=0.
\label{eq-gh}
\eeq
If we put
\beq
\Phi (x) = \exp \left( -i \frac{b_0}{2} (x-x_0)^2 \right) f(x),
\label{transf}
\eeq
we obtain the following reduced Schroedinger-like equation where no first order term remains:
\beq
\frac{\pa^2 f(x)}{\pa x^2} + \left[ -2 c_0 (x-x_0)+(b_0^2-d_0) (x-x_0)^2 \right] f(x)=0.
\eeq
Solutions of this equation are well-known, and are parabolic-cylinder functions $D$.
Coefficients above are related to the original ones as follows:
\beqnl
b_0 &=& \frac{(\pa_x \pa_k G)|_0}{(\pa^2_k G)|_0},\\
c_0 &=& \frac{(\pa_x G)|_0}{(\pa^2_k G)|_0},\\
d_0 &=& \frac{(\pa^2_x G)|_0}{(\pa^2_k G)|_0}.
\eeqnl

We assume $d_0\ll b_0^2$, in view of the fact that, in dispersive model, we expect 
an efficient pair-production for high values of first $x$-derivatives on $n(x)$ 
(or of $v(x)$ in the fluid models; see below), much higher than second $x$-derivatives. [To be 
more precise, it should be also not so strong to induce a coupling with a fourth state B belonging to 
a monotone further branch of the dispersion relation (cf. e.g. \cite{coutant-analytic} and see also Figure \ref{fig:sellm})].
Then we get
\beq
\frac{\pa^2 f(x)}{\pa x^2}   + \left[ -2 c_0 (x-x_0)+b_0^2 (x-x_0)^2 \right] f(x)=0, 
\eeq
which can easily be reduced to the following form: 
\beq
\frac{\pa^2 f(y)}{\pa y^2} + \frac{1}{b_0^2}\left[ y^2 -\frac{c_0^2}{b_0^2}\right] f(y)=0,
\label{parab}
\eeq
which is equivalent to the Schroedinger equation with energy $E=0$ in presence of a 
parabolic potential barrier. We face with a problem: we started with a situation 
where a single group horizon (turning point) was present, and we actually found 
out an effective potential displaying two turning points: the group horizon 
$x_0$ and also 
\beq
x_1 = x_0 + 2 \frac{c_0}{b_0^2}, 
\eeq 
which arises as an effect of the term $O\left((x-x_0)^2\right)$ arising from the transformation 
(\ref{transf}). The latter term should be considered as spurious, higher order, 
as well as $x_1$. Still, in our problem we associate a (small) scale $\epsilon_x$ to each power 
of $(x-x_0)$ and also a (big) scale $L_x$ to the first derivative of $n(x)$. This big scale 
is fundamental for the efficiency of the pair-creation process, and also it violates the 
hypothesis of weakly varying medium occurring e.g. in plasma physics. Then, the term 
$\propto b_0^2$ would be $O(\epsilon_x^2 L_x^2)$, whereas the term $\propto c_0$ 
would be $O(\epsilon_x L_x)$ and the term $\propto d_0$ would be only $O(\epsilon_x^2)$. 
E.g. for $L_x$ order of $\epsilon_x^{-2}$, one would obtain 
$x_1 = x_0+O( \epsilon_x)$, and then $x_1$ would coalesce with $x_0$  in the 
limit as $\epsilon_x \to 0$: it would appear as a so-called secondary turning point 
\cite{fedoryuk-asymptotic}. Of course, in order to obtain $x_1$ coalescing with $x_0$ as 
$\epsilon_x\to 0$ is sufficient to require that $L_x$ is order of $\epsilon_x^{-(1+a)}$, with $a>0$ (so that 
$x_1 = x_0+O( \epsilon_x^a)$).\\
In looking for an approximate and effective 
description of the scattering process, our ansatz is to keep the term $\propto b_0^2$ and 
also $x_1$, as an effective model description where the further turning point 
$x_1$ is assumed to delimit the interaction region where the phenomenon takes place.\\

{\sl We assume to associate a quantum mechanical level with equation (\ref{parab}).} 
This assumption is nontrivial, and is 
justified only by considering a microscopic quantum model associated with the same dispersion 
relation. We need this underlying quantum level, without which both a quantum interpretation 
of the states and quantum tunneling, as we see below, would be lacking.\\
We take into account the transmission probability rate. One could also provide a complete solution, but a WKB 
approximation, within the above approximation of the dispersion relation, will be 
enough. Then we find the following transmission coefficient:
\beq
{\mathcal T}_{WKB}=\exp \left(-\pi \frac{c_0^2}{|b_0|^3}\right).
\label{twkb}
\eeq
This transmission coefficient has to be interpreted as $\Gamma$ \cite{parikh-prl} in the 
tunneling approach to Hawking radiation for black holes. Thus, it is related to 
pair-creation.\\ 
We stress that, in order to relate the above ${\mathcal T}_{WKB}$ to the Hawking effect, 
we should expect the emergence of a dependence on a overall factor  $\omega$, which is  
a non-trivial requirement. Indeed,  
a further step is required, i.e. the possibility to achieve the following identification: 
\beq
\exp \left(-\pi \frac{c_0^2}{|b_0|^3}\right) = \exp \left(-\beta_H \hbar \omega\right),
\eeq
where $\beta_H$ is proportional to the inverse of the Hawking temperature $T_H$. This identification of 
the tunneling probability rate with a Boltzmann factor 
is common in the tunneling approach to the Hawking effect (see e.g. \cite{visser-essentials}). 
We apply our general picture to specific cases in the following.

\subsection{Optical black hole in the Cauchy approximation}
\label{gh-cau}

\begin{figure}[t]
\includegraphics[angle=0,width=8cm]{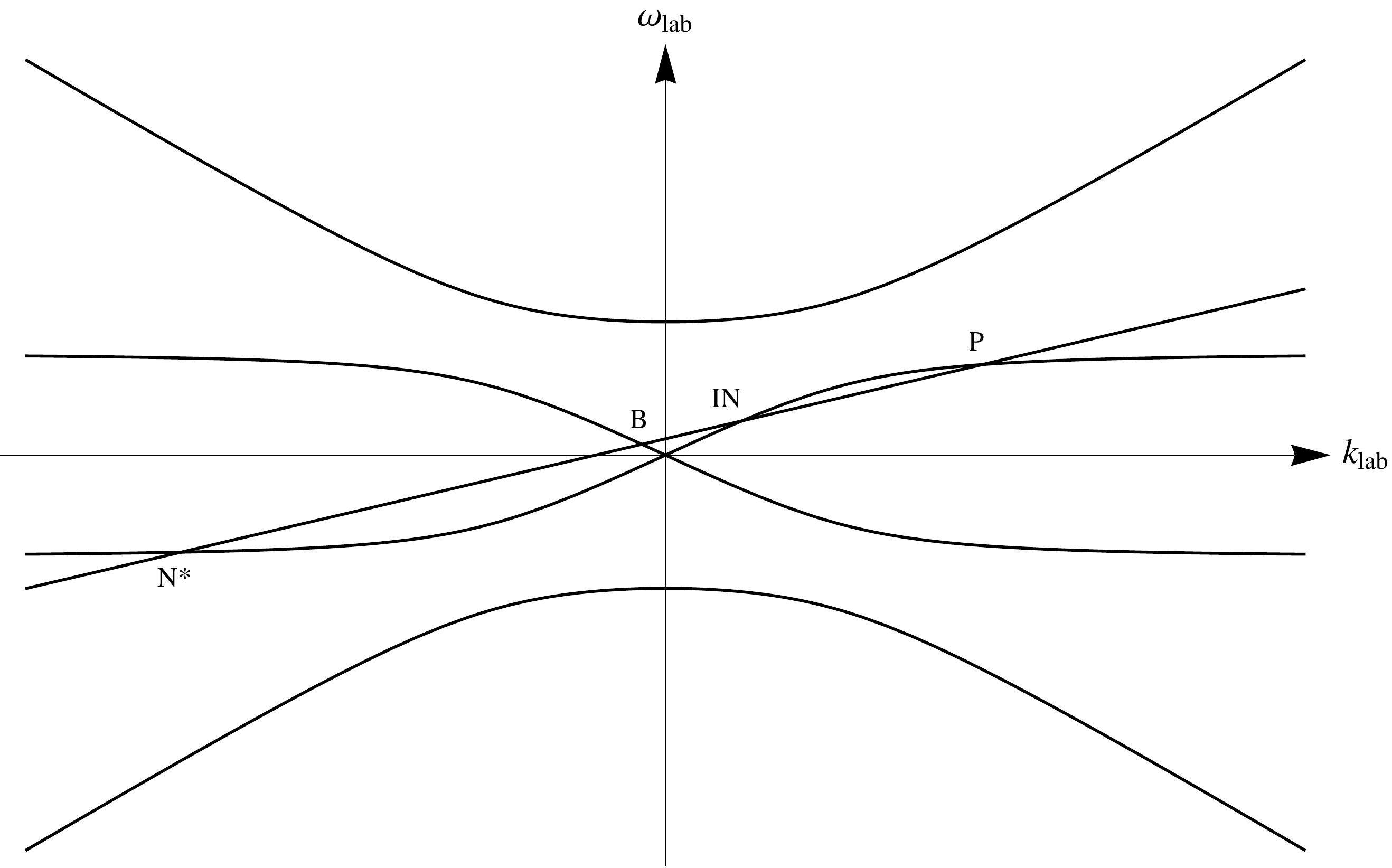}
	\caption{ \label{fig:sellm} Complete asymptotic dispersion relation for the Sellmaier dispersion relation of a diamond-like material in the lab frame (qualitative plot). 
A line of constant $\omega$ is represented by a straight line. Note that a fourth state B appears, which is not actually present if the gradient of the refractive index is not too strong. We are substantially considering only the branch intersected  by the straight line in states IN, P, N$^\ast$.}
\end{figure}

In the case of the Cauchy approximation, i.e. for (\ref{di-cau}), we have
\beqnl
b_0 &=& v (\pa_x n)_0 \frac{(2 B \gamma^4)^{1/3}}{6 v^2 \gamma^2 B} \left(\frac{v}{c}\right)^{1/3} \omega^{-1/3},\\
c_0 &=& (\pa_x n)_0 \frac{1}{6 v^2 \gamma^2 B} ,\\
d_0 &=& (\pa^2_x n)_0 \frac{1}{6 v^2 \gamma^2 B}.
\eeqnl
We find also 
\beq
x_1-x_0=3 2^{4/3} \gamma^{-2/3} \left(\frac{c}{v}\right)^{2/3} (B \omega^2)^{1/3} \frac{1}{|\pa_x n|_0},
\eeq
and 
\beq
{\mathcal T}_{WKB}=\exp \left(-\pi \frac{3 c}{\gamma^2 v^2 |\pa_x n|_0} \omega\right).
\eeq
So we have gained an important corroboration to our approach, because we have found 
the overall factor $\omega$ which was not so trivial to obtain.
Then we find in our approximation
\beq
k_b T_H \sim \frac{\hbar}{c} \frac{\gamma^2 v^2 |\pa_x n|_0}{3 \pi},
\eeq 
which, apart from the factor $3$ in place of $2$ at the denominator, is just the 
expected result.  The missing factor is $1.5$, which could be considered not so bad, given the 
approximation we adopted. Note also that $|\pa_x n|_0$, and then also $T_H$, depend on $\omega$ 
through $x_{GH}(\omega)$. So thermality holds only approximately, when the latter dependence is 
weak. The most direct analogy is with the dependence on $\omega$ of the temperature in the 
Parikh-Wilczek approach. Note that $T_H$ is naturally involving 
`backreaction', in the sense that there is interplay between the external field (refractive index) and the physical field 
(the electromagnetic field) through dispersion, which makes $T_H$ dependent also on $\omega$. It is a 
backreaction in a different sense with respect to the Parikh-Wilczek case, because there is not yet an energy balance for 
the total system `black hole+radiation', but effects of dispersion, which influence the emitted radiation, can be as well 
considered backreaction, being due to an effective interaction between the radiation and the black hole which affects 
the effective description of the black hole itself.\\
We have to clarify a few very important facts: by looking at the dispersion relation which holds 
asymptotically in $x$ (i.e. for $x\to \pm \infty$, where homogeneity is recovered), 
one finds three states involved in the scattering process when a group horizon is present: 
the incident mode $IN$, the positive norm reflected mode $P$, and the negative norm one
$N^\ast$. By looking at the dispersion curve, the group velocity $v_g$ associated with 
$IN$ is positive, %as its phase velocity, 
whereas the group velocities of both $P$ and $N^\ast$ 
are negative in the comoving frame, cf. also Figure \ref{fig:sellm}.
Still, according to Feynman-Stueckelberg interpretation, a negative 
norm state is propagating backward in time, so it is natural to assign to it a 
group velocity which is the opposite of the one which can be deduced from the 
dispersion relation curve. As such, the antiparticle state is propagating 
forward, so it emerges as a transmitted state. The corresponding particle state $N$, 
which actually enters the experimental situation, is propagating backward (i.e. 
is reflected; this is natural in a particle-hole picture, inspired by the Dirac sea picture, where 
the propagating hole is associated with a counterpropagating particle). 
This picture allows us to give concrete meaning to a scattering 
process with a transmitted state which, otherwise, would not be allowed. 
Notice that: a transmitted state in a scattering process is allowable only in  
a quantum mechanical process; moreover, we need an underlying quantum 
field theory model (in our case: Hopfield model) with a conserved norm for 
defining particle and antiparticle states. \\
% Still, the direction of the wave propagation is given by the sign of the product $v_g 
%v_{ph}$, so it is forward for $IN$ and backward for $P$ (which both have positive $v_{ph}$), 
%and is forward for $N^\ast$, which has negative $v_{ph}$. The presence of the 
%negative norm state  $N^\ast$ signals the presence of pair creation. 
% Still, the direction of the wave propagation is given by the sign of the product $v_g 
%v_{ph}$, so it is forward for $IN$ and backward for $P$ (which both have positive $v_{ph}$), 
%and is forward for $N^\ast$, which has negative $v_{ph}$. The presence of the 
%negative norm state  $N^\ast$ signals the presence of pair creation. 
We also point out that we could legitimately solve eqn. (\ref{eq-gh}) and find again 
parabolic-cylinder functions as solutions for our problem, apart from a space-dependent 
phase factor. Calculations of the transmission coefficient ${\mathcal T}_{WKB}$ 
would give the same value as above, and the problem would be to explain 
why ${\mathcal T}_{WKB}>0$ in this case. As a matter of facts, this result 
would not be so unexplainable, as we are dealing with quantum field theory 
in external field, and unexpected transmission coefficients different from 
zero are usual in all situations where e.g. Klein paradox occurs (see e.g. 
\cite{manogue}).\\
As well known, dispersion tends to distort the pure thermality of the non-dispersive 
situation, because at the best one is able to find a temperature which depends on $\omega$. 
Of course, this makes the spectrum not really Planckian. Usually, in the case of weak dispersion, 
one assumes and verifies that, on restricted frequency intervals, it is true that 
$T(\omega)\sim T(\omega=0)=T_H^{nd}$, where $T_H^{nd}$ stays for the non-dispersive temperature.

\subsection{Coutant-Parentani-Finazzi model}
\label{gh-cpf}

As a matter of facts, doubts should be raised on the actual viability of this 
effective approach to 
analogous Hawking radiation in dispersive media. A positive test on a single model could 
be a coincidence. So we try to verify the approach on a different model. 
We start from Coutant-Parentani-Finazzi model \cite{coutant-analytic}, because, at least for a particular 
form for the dispersive contribution, it is possible to carry 
out explicit analytical calculations. Other models could be also checked, 
but, except for the model in the following subsection, 
the system $G=0,\pa_k G=0$ in other cases is quite tricky, so a check seems to be 
difficult.\\
The general class of models considered in \cite{coutant-analytic} satisfies 
\beq
G=(\omega -v(x) k)^2 -F^2 (k)=0,
\label{gen-class}
\eeq
where $F^2 (k)$ is such that dispersive effects are involved. As in 
\cite{coutant-analytic}, we consider the following sub-case:
\beq
G_c:=\omega -v(x) k-\left(c k +\frac{k^3}{2\Lambda^2}\right)=0,
\eeq
where $\Lambda\to \infty$ is the limit as the model becomes non-dispersive and where 
we restored momentarily $c$ (to be identified as $c_{sound}$). We put $c=1$ henceforth. One 
obtains \cite{coutant-analytic} 
\beq
k|_0=-(\omega \Lambda^2)^{1/3},
\eeq
and
\beq
v|_0=-1-\frac{3}{2} \left(\frac{\omega}{\Lambda}\right)^{2/3}.
\eeq
Notice that, as $\Lambda \to \infty$, the group horizon tends to the geometrical horizon 
such that $v|_{geom}=-1$.\\
We obtain 
\beqnl
b_0&=& \frac{\Lambda^2}{3 (\omega \Lambda^2)^{1/3}} (\pa_x v)_0,\\
c_0 &=& \frac{\Lambda^2}{3} (\pa_x v)_0.
\eeqnl
As a consequence, 
\beq
|x_1-x_0|=\frac{6}{\Lambda^{2/3}} \frac{1}{|(\pa_x v)|_0} \omega^{2/3}.
\eeq
From
\beq
{\mathcal T}_{WKB}=\exp \left(-\pi \frac{3 }{|(\pa_x v)|_0} \omega\right),
\eeq
we find in our approximation
\beq
k_b T_H \sim \hbar \frac{|(\pa_x v)|_0}{3 \pi}.
\eeq 
%where we restored $c$ (to be identified as $c_{sound}$). 
Thus the same 
error for a factor $1.5$ as in the optical case is discovered. The temperature is proportional to $(\pa_x v)|_0$, 
as it happens in the corresponding non dispersive case. There is, also in this case, a dependence on $\omega$ 
of the temperature which can be made explicit by choosing a specific velocity profile $v(x)$.\\ 
It is remarkable that the dependence of 
$\omega$ of both $b_0$ and $c_0$ is the same as in the optical model discussed in the previous subsection. 
For what we can ascertain, this seems to be an accident (we start from very different dispersion relations 
and dispersion mechanisms). As discussed in the following, their only apparent link is represented by the 
fact that both models are involved with weak dispersive contributions.\\
In the following subsection, we investigate a further model which can be analytically 
treated. We will find a very different behaviour in $\omega$ for both $b_0$ and $c_0$, but still the overall 
$\omega$ factor appears.

\subsection{Rousseax model}

We take into 
consideration Rousseaux model \cite{rousseaux-prl}, where explicit calculations 
are possible. This model falls in the general class (\ref{gen-class}), but in the 
dispersion relation we are going to write we cannot find a non-dispersive limit 
as before. 
%$\Lambda\to \infty$.\\
The approximate dispersion relation is \cite{rousseaux-prl}
\beq
G=(\omega -v(x) k) -g |k|^{1/2}=0.
\eeq
%We can also find $k(\omega,x)$, 
Assuming for simplicity $k>0$ (results for $k<0$ 
correspond to the case of positive velocity $v$), we can find 
that the equations $G=0$, $\pa_k G=0$ can be solved to find
\beq
v(x_{GH})=-\frac{g}{4\omega},
\eeq
and
\beq
k_0=\frac{4\omega^2}{g}.
\eeq
The group horizon  
exists for $\omega <\left| \frac{g}{4 v}\right|$.\\
We can adopt the same approximation for $G$ around the turning point as above. In 
particular, with reference to equation (\ref{parab}). We have
\beqnl
b_0&=& -\frac{32}{g^2} (\pa_x v)_0 \omega^3,\\
c_0 &=& -\frac{128}{g^3} (\pa_x v)|_0 \omega^5.
\eeqnl
In this case, we obtain
\beq
|x_1-x_0|=\frac{1}{4} \frac{g}{|(\pa_x v)|_0} \frac{1}{\omega}.
\eeq
Then, from
\beq
\exp \left(-\pi \frac{1}{2 |(\pa_x v)|_0} \omega\right) = \exp \left(-\beta_H \hbar \omega\right),
\eeq
one infer
\beq
k_b T_H=\frac{4 \hbar}{\pi} |(\pa_x v)|_0,
\eeq
thus the numerical coefficient is $8$ times the expected one. Not so good, an enhancement of a factor 
$8$, but, on the other hand, 
maybe not so bad, given that both the model is approximate and also the equation is very 
approximate. It is remarkable that, apart from numerical coefficients, the dependence on gradients 
and on $\omega$ are exactly the right ones. As a consequence, the method seems to be just 
more reliable than one could ever have expected. Moreover, it has to be pointed out that the 
model is essentially dispersive, in the sense that one cannot obtain a non-dispersive limit, and so 
it is not clear if one should obtain exactly the `geometrical result' (we mean, the result inspired by the 
non-dispersive case).\\
A natural question one could put about the fluid model above, is why one should choose the above 
approach instead of exact calculations. The point is that in exact calculations the explicit role 
of the group horizon is often hidden inside nontrivial studies of asymptotic expansions of 
fourth order differential equations, which are by no means an easy stuff (almost twenty years 
of still in progress studies quoted above). \\

In concluding this section, we wish to compare the results obtained in all the models we 
have taken under consideration and discuss them further on. 
Thermality arises because of the tunneling effect 
through the group velocity horizon. The process is most efficient as the gradient of the 
refractive index/the fluid velocity field increases (although it is not the only parameter). 
This is also associated with the distinction between adiabatic (particle number conserving) 
and non-adiabatic process (particle number increasing).  
Thermality is very near the expected result in cases where dispersion can be considered 
as weak (optical case, Coutant-Parentani-Finazzi model).  Control over dispersive contributions 
is given by the coefficient $B$ in the optical case (as $B\to 0$ the non-dispersive case is 
recovered) and by $\frac{1}{\Lambda}$ in the  Coutant-Parentani-Finazzi model (non dispersive 
case is recovered as $\Lambda\to \infty$). Both these limit are to be taken with care, because 
they require a suitable regularization in order to find out a reliable result. This is the 
topic to be discussed in the following section.\\
Rousseaux dispersion model 
is still thermal but not near the weak dispersion limit (and indeed dispersion is not 
weak in that case). In this case, we are not aware of any way to recover a sort of 
weak dispersion limit.\\
Tunneling is involved with the presence of a secondary turning point. As a matter of fact, 
the presence of terms $O(x^2)$ and not simply linear in $x$ is necessary for looking for 
a tunnel effect; such a presence of second order terms in $x$ is non-exclusive of our model. 
Indeed, this presence occurs also in \cite{coutant-analytic}, even if in a form which is not 
so evident. Indeed, when the so-called $p$-representation near the horizon $x=0$ is considered in \cite{coutant-analytic}, 
terms $O(x^2)$ appear (they correspond to terms $\propto \pa_p^2$). Consistency would 
also require that even $v(x)$ should contain terms $O(x^2)$. This would in turn imply 
to neglect higher order terms in $\pa_p$ derivatives (up to the fourth order). In view of this, 
the approximation adopted therein seems to be similar to ours one (in neglecting higher 
order terms). Of course, in \cite{coutant-analytic} a much more complete analysis is performed. But we 
wish to remark this relevant point which maybe does not emerge so easily from 
\cite{coutant-analytic}. A thickness for the horizon is also found as in \cite{Coutant-thick}, albeit 
with a different behaviour (and we feel that the approximation in \cite{coutant-analytic,Coutant-thick} is 
more reliable than ours one).\\
In the following section, we deal with the non-dispersive limit and we show that some subtleties, to be 
handled with care, occur.

\section{How to handle the limit of vanishing dispersion 
in the weak dispersion case, and to restore the role of non-dispersive horizon}
\label{expansion-geom}

Let us consider again the expansion (\ref{exp-G}), but this time at first we focus on the non-dispersive 
horizon in place of the group horizon, by keeping dispersive contributions in the dispersion relation. 
Then we get a further term of the first order in $(k-k_0)$, 
which is proportional to $\partial_k G$. This modifies equation (\ref{eq-gh}), so that we obtain 
%\newpage
\beqnl
&&\left( -\frac{1}{2} (\pa^2_k G)|_0 \pa_x^2 -i (\pa_x \pa_k G)|_0 \frac{1}{2} ((x-x_0) \pa_x+\pa_x (x-x_0)) 
-i (\pa_k G)|_0 \pa_x \right.\cr
&&\left. +
(\pa_x G)|_0 (x-x_0)+\frac{1}{2} (\pa^2_x G)|_0 (x-x_0)^2\right) \Phi(x)=0.
\eeqnl
Then the equation is of the form 
\beq
\left(\pa_x^2 +2 i (b_0 (x-x_0)+e_0) \pa_x +(i b_0-2 c_0 (x-x_0)-d_0 (x-x_0)^2) \right) \Phi (x)=0,
\label{eq-geom}
\eeq
where we have defined
\beq
e_0:= \frac{(\pa_k G)|_0}{(\pa^2_k G)|_0}.
\eeq
If we choose
\beq
\Phi (x) = \exp \left( -i \frac{b_0}{2} (x-x_0)^2 -i e_0 (x-x_0)\right) f(x),
\label{transf-geom}
\eeq
we obtain the following reduced Schroedinger-like equation where no first order term remains:
\beq
\frac{\pa^2 f(x)}{\pa x^2} + \left[ -2 c_0 (x-x_0)+\left(b_0 (x-x_0)+e_0\right)^2 -d_0 (x-x_0)^2\right] f(x)=0.
\eeq
Solutions of this equation are again parabolic-cylinder functions $D$.\\ 
Turning points of the parabolic 
potential are 
\beq
x_\pm = x_0+\frac{1}{b_0^2-d_0} \left[ c_0-b_0 e_0 \pm \sqrt{c_0^2 -2 b_0 e_0 c_0+d_0 e_0^2}\right],
\eeq
and the `thickness' of the interaction region is 
\beq
x_+ - x_- =2 \frac{1}{b_0^2-d_0} \sqrt{c_0^2 -2 b_0 e_0 c_0+d_0 e_0^2}.
\eeq
We have to require, in order to get real solutions, 
\beq
c_0^2 -2 b_0 e_0 c_0+d_0 e_0^2\geq 0.
%c_0 (c_0 -2 b_0 e_0)\geq 0.
\eeq
The following redefinition 
\beq
y:=(b_0^2-d_0) x+e_0 b_0-c_0,
% e_0-b_0 (x-x_0) -\frac{c_0}{b_0}
\eeq
allows us to obtain the following form for the Schroedinger-like equation: 
\beq
\left[\frac{d^2}{dy^2} +\frac{1}{(b_0^2-d_0)^3} \left( y^2 -\frac{c_0^2}{b_0^2}+2\frac{c_0 e_0}{b_0} \right) 
\right] f=0,
\eeq
which can be easily turned into the form of a Weber equation, provided that we define 
$z=a^{-3/4} y$.\\
The tunneling coefficient is given by 
\beq
\mathcal{T} = \exp (- \pi \alpha), 
\eeq
with 
\beq
\alpha := \frac{c_0^2 -2 c_0 e_0 b_0+d_0 e_0^2}{|b_0^2-d_0|^{3/2}}.
\label{alfa}
\eeq
In terms of derivatives of the dispersion relation $G$, we get
\beqnl
\alpha :&=& \frac{1}{|(\pa_x \pa_k G)^2-(\pa_k^2 G)(\pa^2_x G)|^{3/2}} \left[(\pa_x G)^2(\pa_k^2 G)-2 (\pa_x G)(\pa_k G) (\pa_x \pa_k G)\right]\cr
&+& \frac{1}{|(\pa_x \pa_k G)^2-(\pa_k^2 G)(\pa^2_x G)|^{3/2}} (\pa_x^2 G)(\pa_k G)^2.
\eeqnl
We consider the above expansion with coefficients whose expressions are suitably 
regularized: if $x_0$ identifies the geometrical horizon, we regularize it by the 
following shift $x_0\mapsto x_0-\epsilon$ (we are approaching a white hole horizon from the left, 
it is easy to arrange for a black hole horizon approached from the right: $x_{bh}\mapsto 
x_{bh}+\epsilon$). Then we perform our calculations, 
perform the limit as $B\to 0^+$ in the optical case and as $\Lambda \to \infty$ in the Coutant-Parentani-Finazzi 
case, and only then we perform also the limit as $\epsilon \to 0$. 
In our view, this is the correct order in which the aforementioned limits are to be taken. Indeed, 
in order to explore the limit in which dispersive effects vanish, we need to consider still 
regularized quantities (such a limit makes singular some expressions). A regularized quantity $R$  
will be indicated as $R_\epsilon$.

\subsection{The optical case: Cauchy approximation}

We implement our calculations by starting from the same dispersion relation as in 
sec. \ref{gh-cau}, with the difference that we are expanding near the geometrical 
horizon and we are working also having in mind the limit of vanishing dispersive effects. 
I.e., we mean to take the limit as $B\to 0^+$. We know that, in the latter limit, 
singularities arise in the wave vector $k$, and then, we adopt the strategy sketched above: 
we first regularize and then let $B\to 0^+$. Only at the end we relax the regularization. 

It is straightforward to show that
\beq
\alpha_\epsilon = 2 \frac{\omega + v k_\epsilon}{v |\pa_x n|_\epsilon} \left( \frac{c}{v}-n\right)_\epsilon
\eeq
holds true; it also holds
\beq
\lim_{B\to 0^+} (\omega + v k_\epsilon) \left( \frac{c}{v}-n\right)_\epsilon = \frac{c}{v} \frac{\omega}{\gamma^2}, 
\eeq
so that 
\beq
\lim_{\epsilon \to 0} \left( \lim_{B\to 0^+} \pi \alpha_\epsilon\right)  = \frac{2 \pi c}{\gamma^2 v^2 |\pa_x n|_0} \omega , 
\eeq
which is the expected result. This is a very intriguing result, as it provides us a further version of the tunneling method 
in which a true effective barrier is to be overcome in order to obtain pair creation. Furthermore, quite surprisingly, 
one can obtain the non-dispersive result from a method devised for the dispersive case. It is also to be noted that, 
in the non-dispersive limit, the thickness of the barrier tends to zero, and one is left with a sort of `phantom barrier' 
in that limit, tending to the geometrical horizon.\\
We also know that (\ref{x-ghgeom}) holds true, i.e. the group horizon shiftes to the geometrical horizon as 
$B\to 0^+$. Then, we can consistently also proceed as follows. As we are interested in the limit as $B\to 0^+$, 
and we know that in such a limit $x_{gh}\to x_{geom}$, also the expansion around the group horizon has to be 
regularized. Our choice is to shift
\beq
x_{gh}\mapsto x_{gh}-\epsilon,
\eeq
in such a way that $x_{gh}-\epsilon \to x_{geom}-\epsilon$ as  $B\to 0^+$. But such a shift can be coherently 
taken into account only at the price to restore the term $(\pa_k G)$, because it is no more valued at $x_{gh}$, 
where it vanishes, but at $x_{gh}-\epsilon$, where it is different from zero (albeit small). Formally,  (\ref{alfa}) 
holds true again. As the limit $B\to 0^+$ is consistently taken before the regularization is relaxed, 
and in such a limit $x_{gh}-\epsilon \to x_{geom}-\epsilon$, we obtain the same result as above. So, we recover the 
expected result even by starting from the group horizon, provided we understand that a regularization procedure 
is necessary. A regularization procedure implying a shift from the horizon coordinate is by no means a novel feature 
of our model: by quoting only the latest approach for deriving Hawking radiation, i.e. the (gravitational 
and gauge) anomaly approach 
introduced by \cite{robinson-wilczek} and then improved by \cite{iso-wilczek}, a regularization is required 
(in \cite{iso-wilczek}, in particular, one has $r_+\mapsto r_+ +\epsilon$, where $r_+$ is the black hole horizon). 
It is also easily shown that, if a regularization is still introduced in the above sense, but no limit of 
vanishing dispersion is taken, when the regularization is sent to zero ($\epsilon \to 0$) the same result as for 
the `on shell' calculation (unregularized) of section \ref{group-hor} is consistently obtained. So, putting 
quantities `on shell' is equivalent to regularizing, computing the physical quantities and then relaxing the 
regularization if dispersive effects are non-vanishing: $B>0$, and $\Lambda <\infty$.

\subsection{The Coutant-Parentani-Finazzi model}

In this case, by taking into account the regularization procedure devised in the previous subsection, we find 
\beq
\alpha_\epsilon = 2 \frac{k_\epsilon}{|\pa_x v|_\epsilon} (v+c )_\epsilon;
\eeq
in the limit as $\Lambda\to \infty$ one has 
\beq
k_\epsilon = \frac{\omega}{v+c},
\eeq
so that 
\beq
\lim_{\epsilon \to 0} \left( \lim_{\Lambda\to \infty} \pi \alpha_\epsilon\right)  = \frac{2 \pi }{|\pa_x v|_0} \omega , 
\eeq
which, again is the correct result for the non-dispersive case.

\section{A discontinuity in the temperature as $B\to 0^+$ ($\Lambda\to \infty$)}
\label{disc-temp}

We face with the main drawback of our model. If we want to explore the limit $B\to 0^+$ for 
weak dispersion, as we are implicitly approaching the geometrical black hole, where, in 
absence of dispersion, we know that divergences in $k$ appear, we have to prescribe 
a regularization. This is by no means a disease, because also in other approaches 
a stretching of the horizon coordinate, to be removed at the end of the calculations, 
is to be introduced. But there is the following problem: we could as well consider 
$T_H (x_{gh}(\omega),B)$ for $B>0$ and find a thermality which, in the optical model in the 
Cauchy approximation and in the Coutant-Parentani-Finazzi model, is such that 
\beq
T_H (x_{gh}(\omega),B)\equiv \frac{3}{2} T_H^{nd},
\eeq
where %$T_H$ is the temperature of the non-dispersive case and 
we use $\equiv$ for indicating that the overall functional dependences are the same 
in the two cases and `nd' stands for non dispersive case. Still, if we wish to explore the non-dispersive limit, we have to 
regularize again also the group horizon coordinate and find the correct result. 
What happens is that
\beq
\lim_{\epsilon \to 0} \left( \lim_{B\to 0^+} T_H (x_{gh}(\omega)-\epsilon,B) \right) 
\not = \lim_{B\to 0^+} \left( \lim_{\epsilon \to 0}  T_H (x_{gh}(\omega)-\epsilon,B) \right).
\eeq
The temperature is not continuous in $(x_{geom},0)$. Even if the regularization 
prescription works well and can be used as a prescription for calculationg the 
non-dispersive result in a new way, from a physical point of view there is a 
unsatisfactory situation, which is even worst if one focuses on the geometrical 
horizon at $B>0$. Indeed, there, when $d_0=0$, the temperature diverges, and 
thermality is still lost if $d_0>0$: 
\beq
\alpha (x_{geom},B) \sim \frac{9 \gamma^4 B^2 (\omega + v k)^5 \pa_x^2 n}{v |\pa_x n|^3},
\eeq
where we have neglected terms proportional to $\pa_x^2 n$ with respect to the 
ones proportional to $\pa_x n$ (or integer powers of the latter). This drawback 
is non-eliminable. If thermality were to be defined only as far as dispersion is 
very weak, one could assume that  
\beq
\lim_{\epsilon \to 0} \left( \lim_{B\to 0^+} T_H (x_{gh}(\omega)-\epsilon,B) \right) 
\eeq
would be a good definition. One could also recover a sort of `continuity along curves', 
in the sense that, if we require that $\epsilon$ is a generic continuous function of $B$, with the 
only requirement to vanish in the limit as $B\to 0^+$, we would still find 
\beq
\lim_{B\to 0^+} T_H (x_{gh}(\omega)-\epsilon(B),B) =T_H^{nd}. 
\eeq
Still, the latter definition of thermality we feel that is not appropriate, 
and a dichotomic behavior remains.

\section{A further point to recover thermality with a continuous nondispersive limit}
\label{inn-hor}

Let us consider a further hypothesis: pair creation happens at a different point $x_\ast$ 
such that a) there is thermality up to corrections which vanish as $B\to 0^+$; b) the 
point $x_\ast$ also tends to $x_{geom}$ in the non-dispersive limit. We start by discussing the 
optical case. Assumption (a) is really strong, but it is corroborated by the fact that 
the right thermality is found in the non-dispersive limit, and, moreover, by the idea that 
there should be a continuous behaviour of the temperature in the same limit.  
It is straightforward to show that, again, whichever explicit expression one can gain at a 
specific point, it holds
\beq
\pi \alpha = \frac{2 \pi }{v |\pa_x n|_\ast} \left(\omega + v k\right)_\ast \left( \frac{c}{v}-n\right)_\ast,
\label{onshell}
\eeq
where all quantities are `on shell' (i.e. satisfy the dispersion relation) at $x=x_\ast$. Then, we require that 
\beq
\left(\omega + v k\right)_\ast \left( \frac{c}{v}-n\right)_\ast = \frac{c}{v} \frac{1}{\gamma^2} \omega (1+a(B) \omega^z),
\label{thermo-ast}
\eeq
where $z$ is a fixed non-negative real number and 
\beq
\lim_{B\to 0^+} a(B) = 0.
\eeq
Of course, we could replace the right hand side of (\ref{thermo-ast}) with a series, where a number of 
coefficients $a_i (B)$ appear. But we assume to consider only the lowest order correction to the thermal 
contribution. Notice that the presence of a correction $a(B) \omega^{z+1}$ is mandatory in order to 
be able to obtain a point where the spectrum is thermal (if $a(B)=0$, then there is no solution).\\
Substituting in the dispersion relation (\ref{di-cau}), one finds 
\beqnl
\left(\omega + v k\right)_\ast &=& \left( a(B) \frac{1}{B \gamma^4} \frac{c}{v} \right)^{1/3} \omega^{(1+z)/3},\\
\left( \frac{c}{v}-n\right)_\ast  &=& \left(\frac{B \gamma^4}{a(B)}\right)^{1/3} 
\left(\frac{c}{v} \right)^{2/3} \frac{1}{\gamma^2} \omega^{(2-z)/3} (1+a(B) \omega^z). \label{cv-ast}
\eeqnl
As we want that $x_\ast \to x_{geom}$ as $B\to 0^+$, we also require that 
\beq
\lim_{B\to 0^+} \frac{B}{a(B)} = 0.
\eeq
From the ratio 
\beq
R:= \frac{ \left( \frac{c}{v}-n\right)_\ast}{ \left( \frac{c}{v}-n\right)_{gh}},
\eeq
recalling the hypothesis $\pa_x n>0$, it is easy to show that, for fixed $\omega$, 
if $a (B)>0$, there is no 
possiblity to get $0<R<1$ in a right neighbourhood of $B=0$, i.e. one finds that 
$x_\ast (\omega)<x_{gh} (\omega) <x_{geom}$. If, instead, $a (B)<0$, one finds that 
$x_{gh} (\omega) <x_{geom}<x_\ast (\omega)$, i.e. the points falls beyond the geometrical 
horizon. The latter case, on the grounds of (\ref{onshell}), appears to be more appealing 
under the hypothesis of positive gradient, because it would maximize the temperature, and would 
imply a partial wave penetration beyond both the group horizon and also the geometrical 
horizon, in a region of stronger gradient. We consider the latter hypothesis as  a bit more appealing, 
and we indicate $x_\ast$ as `inner horizon'. We remark also that, for fixed $B$, it is instead possible 
to find $R<1$ and then it is possible to get a point $x_\ast (\omega)$ coinciding with a 
group horizon $x_{gh}(\bar{\omega})$ for $\bar{\omega}<\omega$.\\  
As to the limit as $B\to 0^+$, the aforementioned overlap does not occur, and what happens is that 
the existence of $x_\ast (\omega)$ implies the existence of a corresponding $x_{gh}(\omega)$, 
whereas the vice-versa is not true in general, i.e. it may happen that a group horizon does 
not imply the existence of $x_\ast$ at the same $\omega$. 
A continuous behaviour of the temperature would emerge, with 
the correct result in the non-dispersive limit without explicitly requiring any regularization. 
The point is that, in this case, the two limits $\epsilon \to 0$ and $B\to 0^+$ commute.   
As a consequence, taking seriously the 
hypothesis of pair-creation at $x_\ast$, it seems that the presence of a group horizon 
is a necessary but not sufficient condition for thermality with the expected temperature. 
But we stress that, as emerges from the above discussion, the point $x_\ast$ is constructed 
`ad hoc'.\\

The same construction can be devised for the Coutant-Parentani-Finazzi model. We sketch the 
main results:
\beqnl
k_\ast &=& \left(- a(\Lambda) 2 \Lambda^2 \right)^{1/3} \omega^{(1+z)/3},\\
\left( v+c\right)_\ast  &=& \left(\frac{1}{-a(\Lambda) 2 \Lambda^2 }\right)^{1/3}  
\omega^{(2-z)/3} (1+a(\Lambda) \omega^z). \label{v-ast}
\eeqnl 
We have to require %both $\lim_{\lambda \to \infty} a(\Lambda)=0$ and also  
$\lim_{\Lambda\to  \infty} a(\Lambda) \Lambda^2 =0$, in such a way that $x_\ast (\omega,\Lambda) \to x_{geom}$ 
as $\Lambda\to  \infty$. Also in this case, for $a(\Lambda)>0$ and for $\pa_x v>0$, one finds 
$x_\ast (\omega)<x_{gh}(\omega)<x_{geom}$; instead, for $a(\Lambda)<0$, one finds 
$x_{gh} (\omega) <x_{geom}<x_\ast (\omega)$. 
This would appear as compatible with the idea of `horizon thickness' as depicted in \cite{Coutant-thick}.

\section{Conclusions}
\label{concl}

We have proposed an effective scattering model for the Hawking effect in analogue black holes, 
inferred from an expansion of the dispersion relation near the group horizon 
(turning point). A Schroedinger-like equation in presence of a suitable parabolic 
barrier is obtained. Computation of the transmission coefficient in the WKB 
approximation gives us a Boltzmann factor which is assumed to be 
a sufficient condition for thermality, as usually in standard tunneling method. 
Changing dispersion relation simply changes 
the value of the coefficients, and so our approach is universally applicable to all 
phenomenological models where an underlying quantum microscopic model is available. It is also remarkable that 
there is no free parameter at hand.\\
We started by studying the group horizon, which is naively the best candidate 
for pair-creation in a dispersive situation. 
The temperature obtained taking into account the group horizon has the
correct dependence on the physical parameters but it can differs from the
correct one just for a numeric factor.  This does not seems a fatal disease of the model (at least as far as 
the optical model and Coutant-Parentani-Finazzi one 
are concerned), because it is reasonably attributed to the approximation itself. 
Nevetheless, with a suitable regularization and completion of the expansion series, 
the correct result can be still restored in the limit as dispersive effects vanish, where 
the group horizon merges with the geometrical horizon. Then a unpleasant dichotomy for 
the behaviour of the temperature appears. 
Indeed, according to our model, there would be a discontinuity in the behavior of the temperature 
in the nondispersive limit; there is a dichotomy between $T_H (x_{gh}; B)$ ($T_H (x_{gh}; 1/\Lambda)$) and $T_H (x_{geom},0)$. 
We have found that regularization allows to obtain the correct convergence of $T_H (x_{gh}; B)$ to 
$T_H (x_{geom},0)$ (which is not true without regularizing). $T_H (x_{gh}; B)$ is not the temperature expected  
(and the measured and/or the calculated one).  We stress that, anyway, one could solve the dichotomy 
by assuming that it arises only because of the approximation involved in the problem: a better model could 
give the correct answer at the group horizon and the correct limit as dispersive effect vanish. In this 
hypothesis, the most natural substitute of the geometrical horizon when dispersive effects 
act, i.e. the group horizon, is the locus where pair creation occurs. This solution would 
be plausible but some more discussion is also viable.\\
By focusing on the geometrical horizon, in the limit of vanishing dispersion, one can recover the 
correct result for the non-dispersive case, by treating with care a process of regularization and 
the sequence of limits to be considered. It is worth pointing out that such a method, which starts 
from a dispersive situation, is interesting in itself, as is able to reproduce standard nondispersive results, and to 
corroborate in a unexpected way the idea of a barrier connected with the pair creation process 
by black holes. In the model, indeed, it appears a parabolic barrier whose turning points converge 
to the geometrical horizon as the dispersive effects are taken to be vanishing. Differently from the 
Parikh-Wilczek approach, where the barrier is self-tuned by backreaction, here it is induced by 
the presence of nonlinear interactions of the fields at hand. So 
our tunneling ansatz, as rough as it can be, still has elements of interest: it seems to 
be universal in the above sense. 
On the other hand, at finite dispersive effects, the serious dichotomy 
problem we found for the group horizon is even worse at the geometrical horizon, 
as discussed in the previous section.\\ 
A possible solution, which has the drawback to be constructed ad hoc, is to attribute 
pair creation to a further point, the `inner horizon' $x_\ast$, which is such that 
it allows to find out thermality up to corrections which vanish in the limit of negligible dispersion.  
According to this hypothesis, thermality with a full continuous behaviour in the 
limit as dispersive effects vanish is possible if pair-creation occurs at a inner horizon $x_\ast (\omega)$, 
which also converges to the geometrical horizon in the limit. Moreover, in the same limit, it implies 
the existence of a group horizon as a necessary (but not sufficient) condition for thermality with the 
above characteristics. Further models and a better analysis are necessary in order to delve into problems 
of the locus or, more probably, the region where particle creation occurs. See \cite{Coutant-thick} 
for recent results.\\
As to the 4D case, we note that, if we allow variable separation also for transverse variables $y,z$ (cf. e.g. the separability 
in $\theta,\phi$ in the spherosymmetrical general relativistic case), there are no substantial modification to the picture 
described above. Indeed, two more conserved quantities $k_y,k_z$ appear in the dispersion relation, which enter in the 
various branches $k_B$ but do not affect the form of the differential equation.

\begin{appendix}

\section{group horizon: from analogous black hole to general relativistic ones}
\label{gh-gr}

In the eikonal approximation and in the non-dispersive case, the vanishing of 
$v_g$ still allows to define the horizon, both in the gravitational and in the analogue cases. 
Let us consider the Schwarzschild case. In Painleve-Gullstrand coordinates, which are regular at the 
horizon, null radial geodesics correspond to the equation 
\beq
\frac{dr}{dt}=\pm 1 - \sqrt{\frac{2M}{r}}=:\dot{r},
\eeq
where the plus sign holds for the outgoing geodesics. Null geodesics can be naturally assumed to 
represent, in geometrical optics approximation, trajectories described by light rays. We can identify 
$v_g=\dot{r}$, and it is evident that group horizon and horizon coincide. Note also that 
\beq
\frac{d v_g}{dr}\Big|_{2M}=\frac{1}{4M},
\eeq
i.e. the gradient of the group velocity at the horizon coincides with the surface gravity.\\

We apply the definition of group horizon to the case of a non-dispersive dielectric black hole. We show that 
we can find the correct horizon condition even without recurring to the notion of analogous metric.  
In the comoving frame, let us consider in the nondispersive case the existence condition for the GH: 
\beq
v_g= \frac{c \left(1-\frac{v}{c} n (x) \right)}{\frac{v}{c} -n(x)}=0.
\eeq
We find the condition 
\beq
n (x)=\frac{c}{v},
\eeq
which is the same found in \cite{belgiorno-prd}. 
It is also interesting to consider the 
following quantity:
\beq
\left(\frac{\partial v_g}{\partial x}\right)_{x_h}=\frac{1}{c} \gamma^2 v^2 (\partial_x n)_{x_h}\equiv \frac{1}{c} \kappa,
\label{surf-grav}
\eeq
where $\kappa$ is the surface gravity associated with the same model as in  \cite{belgiorno-prd}.\\
% Effects that optical dispersion, together with the help coming from wave optics, 
%to be compared with geometrical optics, are effective in taming the divergence. See e.g. \cite{rousseaux-book}. \\

\section{A form of Parikh-Wilczek approach from geometrical optics}
\label{pw-geomopt}

Let us consider a photon which tunnels through a black hole group horizon, by passing 
from inside to outside. In particular, if $x$ is the spatial coordinate in a 2D model, 
we have to find out, in the geometrical optics approximation, the  following quantity: 
\beq
\mathrm{Im}\; \int_{x_{in}}^{x_{out}} dx\; k(x),
\eeq
where $k$ stays for the usual wave vector and $x_{in}$ and $x_{out}$ are coordinates of points 
inside and outside the group horizon.
Indeed, we are interested in the probability of tunneling, which is equivalent to the probability 
of pair creation, and it is given by the following expression:
\beq
\Gamma \sim \exp \left({-2\; \mathrm{Im}\int_{x_{in}}^{x_{out}} dx\; k(x)} \right). 
\eeq
In such an approximation, the effective action is replaced by the above tunneling integral.\\

We adopt the same trick as in the original 
paper by Parikh and Wilczek \cite{parikh-prl}:
\beq
\int_{x_{in}}^{x_{out}} dx\; k(x)= \int_{x_{in}}^{x_{out}} dx \int_0^{k} dk'.
\label{pw-trick}
\eeq
(Note that we are assuming that the branch is connected to $k=0$. This is not true in general, 
but it is easy to allow also more general cases by simple shifts). 
Let us follow even more strictly the tricks in \cite{parikh-prl}. $G$ plays the role of Hamiltonian for geometrical optics, as known \cite{kravtsov}. It is 
also useful, for our purposes, to write
\beq
G = 0 \quad \Longleftrightarrow \quad \Pi_b G_b=0,
\eeq
where the index $b$ indicates the branches of the dispersion relation and, more 
explicitly, 
\beq
G_b = \omega - f_b (k,x).
\eeq
%Then, $\omega$ plays the role of Hamiltonian in the Hamilton-Jacobi equation.  
Then, we take into account that the group velocity (\ref{group-v}) 
on a given branch becomes 
\beq
v_g=\partial_k G_b.
\eeq
From the Hamiltonian equations we obtain 
\beq
\frac{dx}{dt}=v_g=\partial_k G_b
\label{ve-group}
\eeq
and then, by taking into account (\ref{ve-group}) and 
\beq
(dG_b)|_{x,t,\omega}=\partial_k G_b\; dk,
\eeq
we get, for $d\omega' \equiv (dG_b)|_{x,t,\omega}$:
\beq
\int_{x_{in}}^{x_{out}} dx \int_0^{k} dk'=\int_{x_{in}}^{x_{out}} dx \int_0^\omega d\omega' \frac{1}{v_g}.
\label{eq-vg}
\eeq
We exchange the order of integration and then {\sl we assume that $v_g$ is analytical in a neighbourhood 
of the group horizon for each fixed $\omega$, and that $\frac{\partial v_g}{\partial x}\Big|_{x_{GH}} \not =0$}. As a consequence, we can 
consider the contribution to the imaginary part 
of the above integral in $x$ near the group horizon $x_{GH}$, by means of the following series expansion:
\beq
v_g (x) = \frac{\partial v_g}{\partial x}\Big|_{x_{GH}} (x-x_{GH})+O((x-x_{GH})^2).
\eeq
The latter assumption, as a matter of facts, eventually include only the nondispersive case, 
because finiteness of $k$ in the dispersive case is the reason why $v_g$, as a function of 
$x,\omega$, cannot be analytic and has to be nonanalytic and also integrable near the 
group horizon.\\ 
In the non-dispersive case, the `usual' logarithmic divergence, which is at the root of thermality for 
standard (nondispersive) black holes, occurs. 
As to the imaginary part of the 
above effective action, we don't need a complete knowledge of the integral; it is enough
a deformation of the contour in the lower half complex plane, compatibly with the 
request of correct decay at infinity for particles, which, by means of the so-called fractional residue theorem, 
applied to a semicircle encircling the simple pole at hand, leads to the following expression: 
\beq 
\Gamma  \sim 
\exp \left({- \frac{2\pi}{ \frac{\partial v_g}{\partial x} \big|_{x_{GH}}}\; \omega}\right),
\eeq
which, in light of (\ref{surf-grav}), is the correct result.

\end{appendix}

%%%%%%%%%%%%%%%%%%%%%%%%%%%%%%%%%%%%%%%%%%%%%%%%%%%%%%%%%%%%%%%%%%%%%%%%%%%%%%%%%%%%%%%%%%%%%%%%%%

\end{document}